\documentclass[12pt]{article}
\usepackage{graphicx}

\begin{document}
\bibliographystyle{unsrt}
\setlength{\baselineskip}{18pt}
\parindent 24pt

\title{Entanglement Generation and Evolution \\
in Open Quantum Systems}

\author{Aurelian Isar\\
National Institute of Physics and Nuclear Engineering\\
P.O.Box MG-6, Bucharest-Magurele, Romania\\
E-mail: isar@theory.nipne.ro}

\date{}
\maketitle

\begin{abstract}
In the framework of the theory of open systems based on completely
positive quantum dynamical semigroups,
we study the continuous variable entanglement for a system consisting of
two independent harmonic oscillators interacting with a general environment.
We solve the Kossakowski-Lindblad master equation for the time evolution of the considered system
and describe the entanglement in terms of the covariance matrix for an arbitrary Gaussian input state.
Using Peres--Simon necessary and sufficient criterion for separability of two-mode
Gaussian states, we show that for certain values of diffusion and dissipation coefficients
describing the environment, the state keeps for all times its initial type:
separable or entangled. In other cases, entanglement generation, entanglement sudden death or a periodic collapse and revival of entanglement take place. We analyze also the time evolution of the logarithmic negativity, which characterizes the degree of entanglement of the quantum state.
\end{abstract}

\section{Introduction}

The rapid development of the theory of
quantum information has revived the interest
in open quantum systems in connection, on one side, to the decoherence phenomenon and,
on the other side,
to their capacity of generating entanglement in multi-partite systems interacting with their environments.
Quantum entanglement represents the physical resource in
quantum information science which is indispensable for the
description and performance of such tasks like teleportation,
superdense coding, quantum cryptography and quantum computation
\cite{nie}. Therefore the generation, detection and
manipulation of the entanglement continues to be presently a
problem of intense investigation.

When two systems are immersed in an environment, then, besides and at the same
time with the
quantum decoherence, the environment can also generate a quantum entanglement
of the two systems and therefore an additional mechanism to
correlate them \cite{ben2,ben3}. In certain circumstances,
the environment enhances entanglement and in others it
suppresses the entanglement and the state describing the two
systems becomes separable. The structure
and properties of the environment may be such that not only the
two systems become entangled, but also such that a certain
amount of entanglement survives in the asymptotic long-time
regime. The reason is that even if not directly coupled, the
two systems immersed in the same environment can interact
through the environment itself and it depends on how strong
this indirect interaction is with respect to the quantum decoherence,
whether entanglement can be generated at the beginning of the
evolution and, in the case of an affirmative answer, if it can
be maintained for a definite time or it survives indefinitely
in time \cite{ben2}.

In this work we study, in the framework
of the theory of open quantum systems based on completely positive dynamical
semigroups, the dynamics of the continuous variable entanglement for
a subsystem composed of
two identical harmonic oscillators interacting with its environment. We
are interested in discussing the correlation effect of the
environment, therefore we assume that the two systems are
independent, i.e. they do not interact directly. The initial
state of the subsystem is taken of Gaussian form and the
evolution under the quantum dynamical semigroup assures the
preservation in time of the Gaussian form of the state.

The organizing of the paper is as follows. In Sect. 2 the notion of  the quantum
dynamical semigroup is
defined using the concept of a completely positive map. Then we give the general
form of the Kossakowski-Lindblad quantum mechanical master equation describing the evolution
of open quantum systems in the Markovian approximation.
We mention the role of complete positivity in connection with the quantum
entanglement of systems interacting with an external environment.
In Sec. 3 we write and solve the equations of motion in the Heisenberg
picture for two independent harmonic oscillators interacting with a general
environment.
Then, by using the
Peres-Simon necessary and sufficient condition for separability
of two-mode Gaussian states \cite{per,sim}, we investigate in Sec. 4 the
dynamics of entanglement for the considered subsystem. In particular, with the help of the
asymptotic covariance matrix, we determine the behaviour of the entanglement in the limit
of long times. We show that for certain classes of environments
the initial state evolves asymptotically to an equilibrium
state which is entangled, while for other values of the
parameters describing the environment, the entanglement is
suppressed and the asymptotic state is separable.
We analyze also the time evolution of the logarithmic negativity, which characterizes the degree of entanglement of the quantum state. A summary and conclusions are given in Sec. 5.

\section{Axiomatic theory of open quantum systems}

The time evolution of a closed
physical system is given by a dynamical group $U_t,$ uniquely
determined by its generator $H$, which is the Hamiltonian operator of the
system. The action of the dynamical group $U_t$ on any density matrix $\rho$
from the set $\cal D(H)$ of all density matrices in the Hilbert space $\cal H$ of the
quantum system is defined by
\begin{eqnarray}\rho(t)=U_t(\rho)=e^{-{\frac{i}{\hbar}}Ht}\rho e^{{\frac{i}{\hbar}}Ht}
\end{eqnarray}
for all $t\in(-\infty,\infty)$. According to von Neumann,
density operators $\rho\in\cal D(H)$ are trace class (${\rm Tr}~ \rho<\infty$),
self-adjoint ($\rho^\dagger=\rho$), positive ($\rho>0$) operators with
${\rm Tr}~\rho=1$. All these properties are conserved by the time evolution
defined by $U_t$.

In the case of open quantum systems, the time evolution $\Phi_t$ of the density operator
$\rho(t)=\Phi_t(\rho)$ has to preserve the von Neumann conditions for all times.
It follows that $\Phi_t$ must have the following properties:

$(\rm i)~  \Phi_{t}(\lambda_{1}\rho_{1}+\lambda_{2}\rho_{2})=\lambda_{1}
\Phi_{t}(\rho_{1})+\lambda_{2}\Phi_{t}(\rho_{2})$ for $ \lambda_{1},\lambda_{2}
\ge 0,$ $\lambda_{1}+\lambda_2=1,$ i. e. $\Phi_{t}$ must preserve the convex
structure of $\cal D(H),$

$(\rm ii)~  \Phi_{t}(\rho^\dagger)=\Phi_{t}(\rho)^\dagger,$

$(\rm iii)~  \Phi_{t}(\rho)>0,$

$(\rm iv)~  {\rm Tr}~\Phi_{t}(\rho)=1.$

The time evolution $U_{t}$ for closed systems must be a group $U_{t+s}=U_tU_s.$
We have also $U_{0}(\rho)=\rho$ and $U_{t}(\rho)\to \rho$
in the trace norm when $t\to 0$. The dual group $\widetilde U_{t}$ acting
on the
observables $A\in \cal {B(H)},$ i.e. on the bounded operators on $\cal H$,
is given by
\begin{eqnarray}\widetilde U_{t}(A)=e^{{\frac{i}{\hbar}}Ht}Ae^{-{\frac{i}{\hbar}}Ht}.
\end{eqnarray}
Then $\widetilde U_{t}(AB)=\widetilde U_{t}(A)\widetilde U_{t}(B)$ and
$\widetilde U_{t}(I)=I$, where $I$ is the identity operator on $\cal H.$
Also, $\widetilde U_{t}(A)\to A$
ultraweakly when $t\to 0$ and $\widetilde U_{t}$ is an ultraweakly continuous
mapping \cite{ing,dav,lin}. These mappings have a strong positivity property
called complete positivity:
\begin{eqnarray}\sum_{i,j} B_{i}^\dagger\widetilde U_{t}(A_{i}^\dagger A_{j})B_{j}
 \ge 0,~~A_{i},
B_{i}\in \cal {B(H)}.\end{eqnarray}

In the axiomatic approach to the description of the evolution of open quantum systems
\cite{ing,dav,lin}, one supposes that the time evolution $\Phi _{t}$ of open systems
is not very
different from the time evolution of closed systems. The simplest dynamics
$\Phi _{t}$ which introduces a preferred direction in time,
characteristic for dissipative processes, is that in which the group condition
is replaced by the semigroup condition \cite{ing,kos,gor}
\begin{eqnarray}\Phi_{t+s}=\Phi_{t}\Phi_{s},~ t,s\ge 0.\end{eqnarray}
The complete positivity condition has the form:
\begin{eqnarray}\sum_{i,j} B_{i}^\dagger\widetilde \Phi_{t}(A_{i}^\dagger A_{j})B_{j}
 \ge 0,~~A_{i}, B_{i}\in \cal {B(H)},\label{cposo}\end{eqnarray}
where $\widetilde \Phi_{t}$ denotes the dual of $\Phi_{t}$ acting on $\cal B(H)$ and is
defined by the duality condition
\begin{eqnarray}{\rm Tr}(\Phi_{t}(\rho)A)={\rm Tr}(\rho \widetilde \Phi_{t}(A)).
\end{eqnarray}
Then the conditions ${\rm Tr}\Phi_{t}(\rho)=1$ and
$\widetilde \Phi_{t}(I)=I $
are equivalent. Also the conditions
$\widetilde \Phi_{t}(A)\to A $
ultraweakly when $t \to0$ and $\Phi_{t}(\rho)\to \rho$
in the trace norm when $t \to 0,$ are equivalent.
For the semigroups with these properties and with a more weak property of positivity
than Eq. (\ref{cposo}), namely
\begin{eqnarray}A\ge 0 \to \widetilde \Phi_{t}(A) \ge 0,\end{eqnarray}
it is well known that there exists a (generally unbounded) mapping
$\widetilde L$ -- the generator of $\widetilde \Phi_{t},$ and
$\widetilde \Phi_{t}$ is uniquely determined by $\widetilde L.$
The dual generator of the dual semigroup $\Phi_{t}$ is denoted by $L$:
\begin{eqnarray}{\rm Tr}(L(\rho)A)={\rm Tr}(\rho \widetilde L(A)).\end{eqnarray}
The evolution equations by which $L$ and $\widetilde L$ determine uniquely
$\Phi_{t}$ and $\widetilde \Phi_{t}$, respectively, are
given in the Schr\"odinger and Heisenberg picture as
\begin{eqnarray}\frac{d\Phi_{t}(\rho)}{dt}=L(\Phi_{t}(\rho)) \label{meqs}\end{eqnarray}
and
\begin{eqnarray}\frac{d\widetilde \Phi_{t}(A)}{dt}=\widetilde L(\widetilde \Phi_{t}(A)).
\label{meqh}\end{eqnarray}
These equations replace in the case of open systems the von Neumann-Liouville
equations
\begin{eqnarray}\frac{dU_{t}(\rho)}{dt}=-\frac{i}{\hbar}[H,U_{t}(\rho)]\end{eqnarray}
and
\begin{eqnarray}\frac{d\widetilde U_{t}(A)}{dt}=\frac{i}{\hbar}[H,\widetilde U_{t}(A)],
\end{eqnarray} respectively.
For applications, Eqs. (\ref{meqs}) and (\ref{meqh}) are only useful if the detailed
structure of the generator $L(\widetilde L)$ is known and can be related to
the concrete properties of the open systems described by such equations.
For the class of dynamical semigroups which are completely positive and norm
continuous, the generator $\widetilde L$ is bounded. In many applications the generator
is unbounded.

According to Lindblad \cite{lin}, the following argument can
be used to justify the complete positivity of $\widetilde \Phi_{t}$: if the open system
is extended in a trivial way to a larger system described in a Hilbert space
$\cal H\otimes\cal K$ with the time evolution defined by
\begin{eqnarray}\widetilde W_{t}(A\otimes B)=\widetilde \Phi_{t}(A)\otimes B,
~~A\in{\cal B}({\cal H}),~ B\in{\cal B}({\cal K}),\end{eqnarray}
then the positivity of the states of the compound system will be preserved by
$\widetilde W_{t}$ only if $\widetilde \Phi_{t}$ is completely positive. With this
observation a new equivalent definition of the complete positivity is obtained:
$\widetilde \Phi_{t}$ is completely positive if $\widetilde W_{t}$ is positive for
any finite dimensional Hilbert space $\cal K.$ The physical meaning of complete positivity can
mainly be understood in relation to the existence of entangled states, the typical example
being given by a vector state with a singlet-like structure that cannot be written as a tensor
product of vector states. Positivity property guarantees the physical consistency of evolving
states of single systems, while complete positivity prevents inconsistencies in entangled
composite systems and therefore the existence of entangled states makes
the request of complete positivity necessary \cite{ben2}.

A bounded mapping $\widetilde L:\cal {B(H)}\to \cal {B(H)}$ which satisfies
$\widetilde L(I)=0, ~\widetilde L(A^\dagger)=\widetilde L(A)^\dagger$ and
\begin{eqnarray}\widetilde L(A^\dagger A)-\widetilde L(A^\dagger)A-A^\dagger
\widetilde L(A)\ge 0\end{eqnarray}
is called dissipative. The 2-positivity property of the completely positive
mapping $\widetilde \Phi_{t}$:
\begin{eqnarray}\widetilde \Phi_{t}(A^\dagger A)\ge \widetilde \Phi_{t}(A^\dagger)
\widetilde \Phi_{t}(A),\label{cpos2}\end{eqnarray}
with equality at $t=0$, implies that $\widetilde L$ is dissipative. Conversely, the dissipativity of $\widetilde L$ implies
that $\widetilde \Phi_{t}$ is 2-positive. $\widetilde L$ is called completely
dissipative if all trivial extensions of $\widetilde L$ to a compound system described
by $\cal H\otimes\cal K$ with any finite dimensional Hilbert space $\cal K$ are
dissipative. There exists a one-to-one correspondence
between
the completely positive norm continuous semigroups $\widetilde \Phi_{t}$ and
completely dissipative generators $\widetilde L$. The following structural theorem
gives the most general form of a completely dissipative mapping
$\widetilde L$ \cite{lin}.

{\bf Theorem.}
$\widetilde L$ is completely dissipative and ultraweakly
continuous if and only if it is of the form
\begin{eqnarray}\widetilde L(A)=\frac{i}{\hbar}[H,A]+\frac{1}{2\hbar}
\sum_{j} (V_{j}^\dagger[A,V_{j}]+[V_{j}^\dagger,A]V_{j}),\label{genh}\end{eqnarray}
where $V_{j},~ \sum_{j} V_{j}^\dagger V_{j}\in{\cal B(H)}, ~H\in {\cal B(H)}_
{\rm s.a.}$.

The dual generator on the state space (Schr\"odinger picture) is of the form
\begin{eqnarray}L(\rho)=-\frac{i}{\hbar}[H,\rho]+\frac{1}{2\hbar}\sum_{j}([V_{j}\rho,
V_{j}^\dagger]
+[V_{j},\rho V_{j}^\dagger]). \label{gens}\end{eqnarray}
Eqs. (\ref{meqs}) and (\ref{gens}) give the explicit form of the Kossakowski-Lindblad
master equation, which is the most general
time-homogeneous quantum mechanical Markovian master equation with a bounded
Liouville operator \cite{lin,gor,sand,rev}:
\begin{eqnarray}\frac{d\Phi_t(\rho)}{dt}=-\frac{i}{\hbar}[H,\Phi_t(\rho)]+
\frac{1}{2\hbar}\sum_j([V_j\Phi_t(\rho),V_j^\dagger]+[V_j,\Phi_t(\rho)V_j^\dagger]).
\end{eqnarray}

The assumption of a semigroup dynamics is only
applicable in the limit of weak coupling of the subsystem with its
environment, i.e. for long relaxation times \cite{tal}.
We mention that the majority of Markovian master equations found in the
literature are of this form after some rearrangement of terms, even for
unbounded generators.
It is also an empirical fact for many physically interesting situations that
the time evolutions $\Phi_{t}$ drive the system towards a unique final state
$\rho (\infty)=\lim_{t\to \infty} \Phi_{t}(\rho(0))$ for all
$\rho (0)\in \cal D(H)$.

\section{Time evolution of two independent harmonic oscillators interacting with the environment}

We are interested in the dynamics of entanglement in a subsystem composed of two identical non-interacting
(independent) harmonic oscillators in weak interaction with a general
environment, so that their reduced time evolution can be
described by a Markovian, completely positive quantum dynamical
semigroup.
If $ \widetilde \Phi_t $ is the dynamical semigroup describing the irreversible time
evolution of the open quantum system in the Heisenberg picture, then the
Kossakowski-Lindblad master
equation has the following form for an operator A (see Eqs. (\ref{meqh}), (\ref{genh}))
\cite{lin,gor,sand,rev}:
\begin{eqnarray}\frac{d\widetilde\Phi_t(A)}{dt}=
\frac{i}{\hbar}[H,\widetilde\Phi_t(A)]
+\frac{1}{2\hbar}\sum_j(V_j^{\dagger}[\widetilde\Phi_t(A),
V_j]+[V_j^{\dagger},\widetilde\Phi_t(A)]V_j).\label{masteq}\end{eqnarray}
Here, $H$ denotes the Hamiltonian of the open system
and the operators $V_j, V_j^\dagger,$ defined on the Hilbert space of $H,$
represent the interaction of the open system
with the environment. We are interested
in the set of Gaussian states, therefore we introduce such quantum
dynamical semigroups that preserve this set. Consequently $H$ is
taken to be a polynomial of second degree in the coordinates
$x,y$ and momenta $p_x,p_y$ of the two quantum oscillators and
$V_j,V_j^{\dagger}$ are taken polynomials of first degree
in these canonical observables. Then in the linear space
spanned by the coordinates and momenta there exist only four
linearly independent operators $V_{j=1,2,3,4}$ \cite{san}: \begin{eqnarray}
V_j=a_{xj}p_x+a_{yj}p_y+b_{xj}x+b_{yj}y,\end{eqnarray} where
$a_{xj},a_{yj},b_{xj},b_{yj}\in {\bf C}.$
The Hamiltonian $H$ of the two uncoupled identical harmonic
oscillators of mass $m$ and frequency $\omega$ is given by \begin{eqnarray} H=\frac{1}{2m}(p_x^2+p_y^2)+\frac{m\omega^2}{2}(x^2+y^2).\end{eqnarray}

The fact that $\widetilde \Phi_t$ is a dynamical semigroup
implies the positivity of the following matrix formed by the
scalar products of the four vectors $ {\bf a}_x, {\bf b}_x,
{\bf a}_y, {\bf b}_y,$ whose entries are the components $a_{xj},b_{xj},a_{yj},b_{yj},$
respectively:
\begin{eqnarray}\frac{1}{2} \hbar \left(\matrix
{({\bf a}_x {\bf a}_x)&({\bf a}_x {\bf b}_x) &({\bf a}_x
{\bf a}_y)&({\bf a}_x {\bf b}_y) \cr ({\bf b}_x {\bf a}_x)&({\bf
b}_x {\bf b}_x) &({\bf b}_x {\bf a}_y)&({\bf b}_x {\bf b}_y)
\cr ({\bf a}_y {\bf a}_x)&({\bf a}_y {\bf b}_x) &({\bf a}_y
{\bf a}_y)&({\bf a}_y {\bf b}_y) \cr ({\bf b}_y {\bf
a}_x)&({\bf b}_y {\bf b}_x) &({\bf b}_y {\bf a}_y)&({\bf b}_y
{\bf b}_y)}\right).
\end{eqnarray}
Its matrix elements have to be chosen appropriately to suit various physical models
of the environment. For a quite general environment able to induce noise and
damping effects, we take this matrix of the following form, where all
the coefficients $D_{xx}, D_{xp_x},$... and $\lambda$ are real quantities, representing
the diffusion coefficients and, respectively, the dissipation constant:
\begin{eqnarray} \left(\matrix{D_{xx}&- D_{xp_x} - i \hbar\lambda/2&D_{xy}& -
D_{xp_y} \cr - D_{xp_x} + i \hbar\lambda/2&D_{p_x p_x}&-
D_{yp_x}&D_{p_x p_y} \cr D_{xy}&- D_{y p_x}&D_{yy}&- D_{y p_y}
- i \hbar\lambda/2 \cr - D_{xp_y} &D_{p_x p_y}&- D_{yp_y} + i
\hbar\lambda/2&D_{p_y p_y}}\right).\label{coef} \end{eqnarray}
It follows that
the principal minors of this matrix are positive or zero. From
the Cauchy-Schwarz inequality the following relations for the
coefficients defined in Eq. (\ref{coef}) hold (from now on we
put, for simplicity, $\hbar=1$): \begin{eqnarray}
D_{xx}D_{p_xp_x}-D^2_{xp_x}\ge\frac{\lambda^2}{4},~
D_{yy}D_{p_yp_y}-D^2_{yp_y}\ge\frac{\lambda^2}{4},\nonumber\\
D_{xx}D_{yy}-D^2_{xy}\ge0,~
D_{p_xp_x}D_{p_yp_y}-D^2_{p_xp_y}\ge 0, \nonumber \\
D_{xx}D_{p_yp_y}-D^2_{xp_y}\ge 0,~D_{yy}D_{p_xp_x}-D^2_{yp_x}\ge 0.
\label{coefineq}\end{eqnarray}
The matrix of the coefficients (\ref{coef}) can be conveniently
written as ($\rm T$ denotes the transposed matrix)
\begin{eqnarray}
\left(\begin{array}{cc}C_1&C_3\\
 {C_3}^{\rm T} &C_2 \end{array}\right),\label{subm}
\end{eqnarray}
in terms of $2\times 2$ matrices
$C_1={C_1}^\dagger$, $C_2={C_2}^\dagger$ and ${C_3}$. This
decomposition has a direct physical interpretation: the
elements containing the diagonal contributions $C_1$ and $C_2$
represent diffusion and dissipation coefficients corresponding
to the first, respectively the second, system in absence of the
other, while the elements in $C_3$ represent environment
generated couplings between the two oscillators, taken initially independent.

We introduce the following $4\times 4$ bimodal covariance matrix:
\begin{eqnarray}\sigma(t)=\left(\matrix{\sigma_{xx}(t)&\sigma_{xp_x}(t) &\sigma_{xy}(t)&
\sigma_{xp_y}(t)\cr \sigma_{xp_x}(t)&\sigma_{p_xp_x}(t)&\sigma_{yp_x}(t)
&\sigma_{p_xp_y}(t)\cr \sigma_{xy}(t)&\sigma_{yp_x}(t)&\sigma_{yy}(t)
&\sigma_{yp_y}(t)\cr \sigma_{xp_y}(t)&\sigma_{p_xp_y}(t)&\sigma_{yp_y}(t)
&\sigma_{p_yp_y}(t)}\right),\label{covar} \end{eqnarray} with the correlations of operators $A_1$ and
$A_2,$ defined by using the density operator $\rho$ of the initial state of the quantum system, as follows:
\begin{eqnarray}\sigma_{A_1A_2}(t)={1\over 2}{\rm
Tr}(\rho(A_1A_2+A_2A_1)(t))-{\rm Tr}(\rho A_1(t)){\rm Tr}(\rho A_2(t)).\end{eqnarray}
By using Eq. (\ref{masteq}) we obtain by direct calculation the following systems of equations for the quantum correlations of the canonical observables \cite{san}:
\begin{eqnarray}{d \sigma(t)\over
dt} = Y \sigma(t) + \sigma(t) Y^{\rm T}+2 D,\label{vareq}\end{eqnarray} where
\begin{eqnarray} Y=\left(\matrix{ -\lambda&1/m&0 &0\cr -m\omega^2&-\lambda&0&
0\cr 0&0&-\lambda&1/m \cr 0&0&-m\omega^2&-\lambda}\right),\end{eqnarray}
\begin{eqnarray}D=\left(\matrix{
D_{xx}& D_{xp_x} &D_{xy}& D_{xp_y} \cr D_{xp_x}&D_{p_x p_x}&
D_{yp_x}&D_{p_x p_y} \cr D_{xy}& D_{y p_x}&D_{yy}& D_{y p_y}
\cr D_{xp_y} &D_{p_x p_y}& D_{yp_y} &D_{p_y p_y}} \right).\end{eqnarray}
Introducing the notation
$\sigma(\infty)\equiv\lim_{t\to\infty}\sigma(t),$ the time-dependent
solution of Eq. (\ref{vareq}) is given by \cite{san}
\begin{eqnarray}\sigma(t)= M(t)(\sigma(0)-\sigma(\infty)) M^{\rm
T}(t)+\sigma(\infty),\label{covart}\end{eqnarray} where the matrix $M(t)=\exp(Yt)$ has to fulfill
the condition $\lim_{t\to\infty} M(t) = 0.$
In order that this limit exists, $Y$ must only have eigenvalues
with negative real parts. The values at infinity are obtained
from the equation \begin{eqnarray}
Y\sigma(\infty)+\sigma(\infty) Y^{\rm T}=-2 D.\label{covarinf}\end{eqnarray}

\section{Dynamics of entanglement}

The two-mode Gaussian state is entirely specified by its
covariance matrix (\ref{covar}), which is a real,
symmetric and positive matrix with the following block
structure:
\begin{eqnarray}
\sigma(t)=\left(\begin{array}{cc}A&C\\
C^{\rm T}&B \end{array}\right),
\end{eqnarray}
where $A$, $B$ and $C$ are $2\times 2$ matrices. Their entries
are correlations of the canonical operators $x,y,p_x$ and $p_y;$ $A$
and $B$ denote the symmetric covariance matrices for the
individual reduced one-mode states, while the matrix $C$
contains the cross-correlations between modes. The elements of
the covariance matrix depend on $Y$ and $D$ and can be
calculated from Eqs. (\ref{covart}), (\ref{covarinf}). Since the two oscillators are identical, it is natural to consider environments for which the two
diagonal submatrices in Eq. (\ref{subm}) are equal, $C_1=C_2,$ and the matrix $C_3$ is symmetric,
so that in the following we take $D_{xx}=D_{yy},~ D_{xp_x}=D_{yp_y},~
D_{p_xp_x}=D_{p_yp_y},~ D_{xp_y}=D_{yp_x}.$ Then both unimodal covariance
matrices are equal, $A=B,$ and the entanglement matrix $C$ is
symmetric.

\subsection{Time evolution of entanglement}

It is interesting that the general theory of open quantum
systems allows couplings via the environment between uncoupled
oscillators. According to the definitions of the environment
parameters, the diffusion coefficients above can take non-zero values and therefore can simulate an interaction between the uncoupled
oscillators. Consequently, the cross-correlations
between modes can have non-zero values. In this case the Gaussian states with $\det C\ge 0$ are
separable states, but for $\det C <0$ it may be possible that
the states are entangled, as will be shown next.

In order to investigate whether an external environment can actually
entangle the two independent systems, we use the partial
transposition criterion \cite{per,sim}: a state is
entangled if and only if the operation of partial transposition
does not preserve its positivity. For the particular case of Gaussian states, Simon \cite{sim} obtained the
following necessary and sufficient criterion for separability:
$S\ge 0,$ where \begin{eqnarray} S\equiv\det A \det B+(\frac{1}{4} -|\det
C|)^2- {\rm Tr}[AJCJBJC^{\rm T}J]- \frac{1}{4}(\det A+\det B)
\label{sim1}\end{eqnarray} and $J$ is the $2\times 2$ symplectic matrix
\begin{eqnarray}
J=\left(\begin{array}{cc}0&1\\
-1&0\end{array}\right).
\end{eqnarray}

In the following we consider such environment diffusion coefficients, for which \begin{eqnarray}m^2\omega^2D_{xx}=D_{p_xp_x},~D_{xp_x}=0,
~m^2\omega^2D_{xy}=D_{p_xp_y}.\label{envcoe}\end{eqnarray} This corresponds to the case when the asymptotic state is a Gibbs state \cite{rev}.

In order to describe the dynamics of entanglement, we have to analyze the time evolution of the Simon function $S(t)$ (\ref{sim1}). We consider two cases, according to the type of the initial Gaussian state: separable or entangled.

1) To illustrate a possible generation of the entanglement, we represent in Figure 1 the function $S(t)$ versus time $t$ and diffusion coefficient  $D_{xp_y}\equiv d$ for a separable initial Gaussian state with initial correlations $\sigma_{xx}(0)=1,~\sigma_{p_xp_x}(0)=1/2,~\sigma_{xp_x}(0)=0,~
\sigma_{xy}(0)=0,~\sigma_{p_xp_y}(0)=0,~\sigma_{xp_y}(0)=0.$ We notice that, according to Peres-Simon criterion, for relatively small values of the coefficient $d,$ the initial separable state remains separable for all times. For larger values of $d,$ at some finite moment of time, when $S(t)$ becomes negative, the state becomes entangled. In some cases the entanglement is only temporarily generated, that is the state becomes again separable. After that, at a certain moment of time, one can notice again a revival of entanglement. In these cases the generated entangled state remains entangled forever, including the asymptotic final state.

\begin{figure}
%\resizebox{0.66\columnwidth}{!}
{
\includegraphics{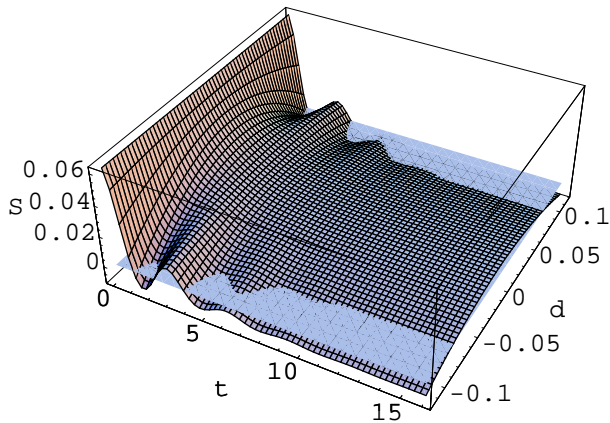}
}
\caption{Simon separability function $S$ versus time $t$
and environment coefficient $d,$ for $\lambda=0.2,$ $D=0.115$ and a separable initial Gaussian state with initial correlations $\sigma_{xx}(0)=1,~\sigma_{p_xp_x}(0)=1/2,~\sigma_{xp_x}(0)=\sigma_{xy}(0)=\sigma_{p_xp_y}(0)=~\sigma_{xp_y}(0)=0.$ We have taken $m=\omega=\hbar=1.$
}
\label{fig:1}
\end{figure}

2) The evolution of an entangled initial state is illustrated in Figure 2, where we represent the function $S(t)$ versus time and diffusion coefficient  $d$ for an initial entangled Gaussian state with initial correlations $\sigma_{xx}(0)=1,~\sigma_{p_xp_x}(0)=1/2,~\sigma_{xp_x}(0)=0,~
\sigma_{xy}(0)=1/2,~\sigma_{p_xp_y}(0)=-1/2,~\sigma_{xp_y}(0)=0.$ We notice that for relatively small values of $d,$ at some finite moment of time $S(t)$ takes non-negative values and therefore the state becomes separable. This is the so-called phenomenon of entanglement sudden death. Depending on the values of the coefficient $d,$ it is also possible to have a repeated collapse and revival of the entanglement. One can also show that for relatively large values of the coefficients $D_{xx}$ and $d,$ the initial entangled state remains entangled for all times.

\begin{figure}
%\resizebox{0.66\columnwidth}{!}
{
\includegraphics{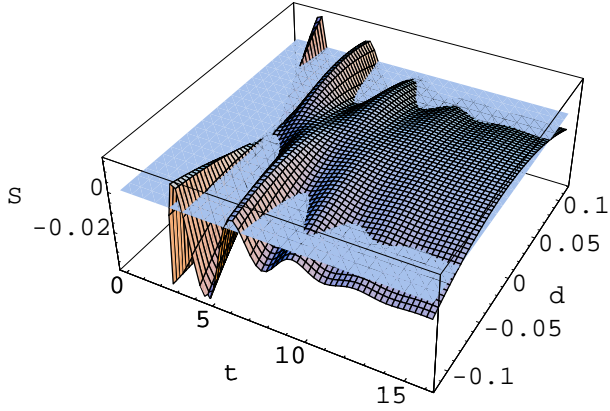}
}
\caption{Same as in Fig. 1, for an entangled initial Gaussian state with initial correlations $\sigma_{xx}(0)=1,~\sigma_{p_xp_x}(0)=1/2,~\sigma_{xp_x}(0)=0,~
\sigma_{xy}(0)=1/2,~\sigma_{p_xp_y}(0)=-1/2,~\sigma_{xp_y}(0)=0.$
}
\label{fig:2}
\end{figure}

\subsection{Asymptotic entanglement}

On general grounds, one expects that the effects of
decoherence, counteracting entanglement production, is dominant
in the long-time regime, so that no quantum correlation (entanglement) is expected to be left at infinity. Nevertheless, there are situations in which the environment allows the
existence of entangled asymptotic equilibrium states.  From Eq. (\ref{covarinf}) we obtain the
following elements of the asymptotic entanglement matrix $C(\infty)$:
\begin{eqnarray}\sigma_{xy} (\infty) =
\frac{m^2(2\lambda^2+\omega^2)D_{xy}+2m\lambda
D_{xp_y}+D_{p_xp_y}} {2m^2\lambda(\lambda^2+\omega^2)},\end{eqnarray}
\begin{eqnarray}\sigma_{xp_y}(\infty)=
\sigma_{yp_x}(\infty)=\frac{-m^2\omega^2 D_{xy}+2m\lambda
D_{xp_y}+ D_{p_xp_y}}{2m(\lambda^2+\omega^2)},\end{eqnarray}
\begin{eqnarray}\sigma_{p_xp_y} (\infty) =
\frac{m^2\omega^4D_{xy}-2m\omega^2\lambda D_{xp_y}+(2\lambda^2
+\omega^2)D_{p_xp_y}}{2\lambda(\lambda^2+\omega^2)}.\end{eqnarray} The
elements of matrices $A(\infty)$ and $B(\infty)$ are obtained by putting $x=y$ in the previous
expressions. We calculate the determinant of the entanglement matrix and obtain:
\begin{eqnarray} \det
C(\infty)~~~~~~~~~~~~~~~~~~~~~~~~~~~~~~~~~~~~~~~~~~~~~~~~~~~~~~~~~~~~~~~~~\nonumber\\
=\frac{1}{4\lambda^2(\lambda^2+\omega^2)}\times[(m\omega^2D_{xy}+
\frac{1}{m}
D_{p_xp_y})^2+4\lambda^2(D_{xy}D_{p_xp_y}-D_{xp_y}^2)].\end{eqnarray}
With the chosen coefficients (\ref{envcoe}), the Simon
expression (\ref{sim1}) takes the following form in the limit of large times: \begin{eqnarray} S(\infty)=
\left(\frac{m^2\omega^2(D_{xx}^2-D_{xy}^2)}{\lambda^2}+
\frac{D_{xp_y}^2}{\lambda^2+\omega^2}-\frac{1}{4}\right)^2-4\frac
{m^2\omega^2D_{xx}^2D_{xp_y}^2}{\lambda^2(\lambda^2+
\omega^2)}.\label{sim2}\end{eqnarray} For environments characterized by
such coefficients that the expression $S(\infty)$ (\ref{sim2}) is strictly negative,
the asymptotic final state is entangled. Just to give an example, without altering the general features of the system, we consider the particular case of $D_{xy}=0.$ Then we obtain that $S(\infty)<0,$ i.e. the asymptotic final
state is entangled, for the following range of values of the
coefficient $D_{xp_y}$ characterizing the environment \cite{arus,aqinf}:
\begin{eqnarray}
\frac{m\omega
D_{xx}}{\lambda}-\frac{1}{2}<\frac{D_{xp_y}}{\sqrt{\lambda^2
+\omega^2}}<\frac{m\omega
D_{xx}}{\lambda}+\frac{1}{2},\label{insep}\end{eqnarray} where the
diffusion coefficient $D_{xx}$ satisfies the condition $m\omega
D_{xx}/\lambda\ge 1/2,$ equivalent with the unimodal
uncertainty relation. We remind that, according to inequalities (\ref{coefineq}), the coefficients have to fulfill also the constraint $D_{xx}\ge D_{xp_y}.$ If the coefficients do not fulfil the
inequalities (\ref{insep}), then $S(\infty)\ge 0$ and the
asymptotic state of the considered system is
separable. These results show that, irrespective of the initial conditions,
we can obtain either an separable or an inseparable asymptotic entangled state,
for a suitable choice of the diffusion and dissipation coefficients.

\subsection{Logarithmic negativity}

\begin{figure}
%\resizebox{0.66\columnwidth}{!}
{
\includegraphics{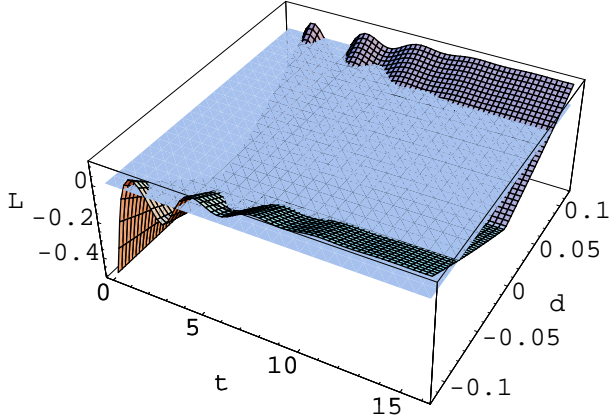}
}
\caption{Logarithmic negativity $L$ versus time $t$
and environment coefficient $d,$ for $\lambda=0.2,$ $D=0.115$ and a separable initial Gaussian state with initial correlations $\sigma_{xx}(0)=1,~\sigma_{p_xp_x}(0)=1/2,~\sigma_{xp_x}(0)=\sigma_{xy}(0)=\sigma_{p_xp_y}(0)=~\sigma_{xp_y}(0)=0.$ We have taken $m=\omega=\hbar=1.$
}
\label{fig:3}
\end{figure}

We apply the measure of entanglement based on negative eigenvalues
of the partial transpose of the subsystem density matrix. For a Gaussian density
operator, the negativity is completely defined
by the symplectic spectrum of the partial transpose of the covariance matrix. The logarithmic negativity $L(t)=-\frac{1}{2}\log_2[4f(\sigma(t))]$
determines the strength of entanglement for $L(t)>0.$ If $L(t)\le 0,$ then the state is
separable. Here \begin{eqnarray} f(\sigma(t))=\frac{1}{2}(\det A +\det
B)-\det C\nonumber\\
-\left({\left[\frac{1}{2}(\det A+\det B)-\det
C\right]^2-\det\sigma(t)}\right)^{1/2}.\end{eqnarray}
In Figures 3 and 4 we represent the logarithmic negativity $L(t)$ versus time $t$ and diffusion coefficient  $D_{xp_y}\equiv d$ for the two types of the initial Gaussian state, separable or entangled, with the same initial correlations, previously considered when we analyzed the time evolution of the Simon function $S(t).$ As expected, we remark that the logarithmic negativity has a behaviour similar to that one of the Simon function in what concerns the characteristics of the state of being separable or entangled. Depending on the values of the environment coefficients, the initial state can preserve for all times its initial property -- separable or entangled, but we can also notice the generation of entanglement or the collapse of entanglement  (entanglement sudden death) at those finite moments of time when the logarithmic negativity $L(t)$ reaches zero value. One can also observe a repeated collapse and revival of the entanglement. In the case of an entangled initial state, the logarithmic negativity is a fluctuating function and decreases asymptotically in time.

\begin{figure}
%\resizebox{0.66\columnwidth}{!}
{
\includegraphics{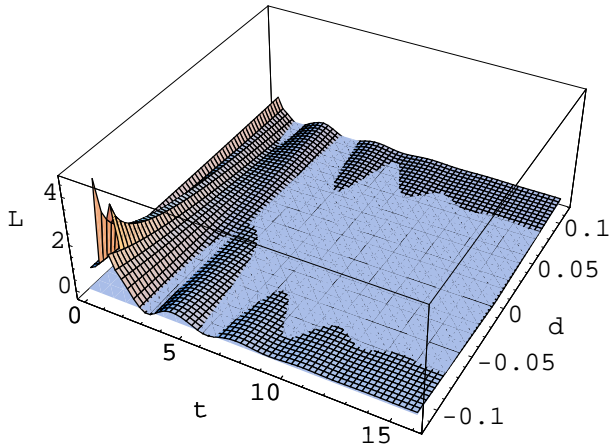}
}
\caption{Same as in Fig. 3, for an entangled initial Gaussian state with initial correlations $\sigma_{xx}(0)=1,~\sigma_{p_xp_x}(0)=1/2,~\sigma_{xp_x}(0)=0,~
\sigma_{xy}(0)=1/2,~\sigma_{p_xp_y}(0)=-1/2,~\sigma_{xp_y}(0)=0.$
}
\label{fig:4}
\end{figure}

In our case the asymptotic logarithmic negativity has the form
\begin{eqnarray} L(\infty)=-\log_2\left[2\left|\frac{m\omega
D_{xx}}{\lambda}-\frac{D_{xp_y}}{\sqrt{\lambda^2
+\omega^2}}\right|\right].\label{aslog}\end{eqnarray}
It depends only on the diffusion and dissipation coefficients
characterizing the environment and does not depend on the initial
Gaussian state. One can easily see that the double inequality (\ref{insep}),
determining the existence of asymptotic entangled states  ($S(\infty)<0$) is equivalent with the condition
of positivity of the expression (\ref{aslog}) of the logarithmic negativity,  $L(\infty)>0.$

\section{Summary}

We have given a brief review of the theory of open quantum systems
based on completely positive quantum dynamical
semigroups and mentioned the necessity of the complete positivity
for the existence of entangled states of systems interacting with an
external environment. In the framework of this theory we investigated
the dynamics of the quantum entanglement for a subsystem
composed of two uncoupled identical harmonic oscillators
interacting with a common environment.

By using the Peres-Simon
necessary and sufficient condition for separability of two-mode
Gaussian states, we have described the generation and evolution of entanglement in terms
of the covariance matrix for an arbitrary Gaussian input
state. For some values of diffusion and dissipation coefficients
describing the environment, the state keeps for all times its initial type:
separable or entangled. In other cases, entanglement generation or entanglement suppression (entanglement sudden death) take place or even one can notice a repeated collapse and revival of entanglement.
We have also shown that, independent of the type of the
initial state, for certain classes of
environments the initial state evolves asymptotically to an
equilibrium state which is entangled, while for other values of the
coefficients describing the environment, the asymptotic state
is separable. We described also the time evolution of the logarithmic negativity, which characterizes the degree of entanglement of the quantum state.

The existence of quantum correlations between the two considered harmonic oscillators interacting with a common environment is the result of the competition between entanglement and quantum decoherence.
From the formal point of view, the generation of entanglement or its suppression (entanglement sudden death) correspond to the finite time vanishing of the Simon separability function or, respectively, of the logarithmic negativity. Presently there is a large debate relative to the physical interpretation existing behind these fascinating phenomena. Due to the increased interest manifested towards
the continuous variables approach  \cite{bra} to quantum information theory,
these results, in particular the
possibility of maintaining a bipartite entanglement in a
diffusive-dissipative environment for asymptotic long
times, might be useful in controlling the entanglement in open systems and also for applications in the field of quantum information
processing and communication.

\section*{Acknowledgments}

The author acknowledges the financial support received within
the Project CEEX 68/2005.


\begin{thebibliography}{99}
\bibitem{nie}
M. A. Nielsen and I. L. Chuang, {\it Quantum Computation and Quantum
Information,} Cambridge University Press, Cambridge 2000.

\bibitem{ben2} F. Benatti and R. Floreanini, {\em Int. J. Mod.
Phys. B} {\bf 19}, 3063 (2005).

\bibitem{ben3}
F. Benatti and R. Floreanini, {\em J. Phys. A: Math. Gen.} {\bf 39},
2689 (2006).

\bibitem{per}
A. Peres, {\em Phys. Rev. Lett.} {\bf 77}, 1413 (1996).

\bibitem{sim}
R. Simon, {\em Phys. Rev. Lett.} {\bf 84}, 2726 (2000).

\bibitem{ing}
R. S. Ingarden and A. Kossakowski, {\em Ann. Phys.} (N.Y.) {\bf 89}, 451 (1975).

\bibitem{dav}
E. B. Davies, {\it Quantum Theory of Open Systems,} Academic Press, New York 1976.

\bibitem{lin}
G. Lindblad, {\em Commun. Math. Phys.} {\bf 48}, 119 (1976).

\bibitem{kos}
A. Kossakowski, {\em Rep. Math. Phys.} {\bf 3}, 247 (1972).

\bibitem{gor}
V. Gorini, A. Kossakowski, E. C. G. Sudarshan, {\em J. Math. Phys.} {\bf 17},
821 (1976).

\bibitem{sand}
A. Sandulescu and H. Scutaru, {\em Ann. Phys.} (N.Y.) {\bf 173}, 277 (1987).

\bibitem{rev}
A. Isar, A. Sandulescu, H. Scutaru, E. Stefanescu and W. Scheid,
{\em Int. J. Mod. Phys. E} {\bf 3}, 635 (1994).

\bibitem{tal}
P. Talkner, {\em Ann. Phys.} (N.Y.) {\bf 167}, 390 (1986).

\bibitem{san}
A. Sandulescu, H. Scutaru, W. Scheid, {\em J. Phys. A: Math. Gen.}  {\bf 20}, 2121 (1987).

\bibitem{arus}
A. Isar, {\it J. Russ. Laser Res.} {\bf 28}, 439 (2007).

\bibitem{aqinf}
A. Isar, {\it Int. J. Quantum Inf.} {\bf 6}, 689 (2008).

\bibitem{bra}
{\it Quantum Information with Continuous Variables}, ed. by
S. L. Braunstein and A. K. Pati, Kluwer, Dordrecht 2003.

\end{thebibliography}
\end{document}